\documentclass[%
 reprint,
aps,
nofootinbib]{revtex4-1}
\usepackage{scrextend}
\usepackage{graphicx}
\usepackage{dcolumn}
\usepackage{bm}
\usepackage{float}
\usepackage[mathlines]{lineno}
\usepackage{footnote}
\usepackage{amsmath}
\usepackage{xfrac}
\usepackage{braket}
\usepackage{cancel}
\usepackage{xcolor}
\usepackage{amssymb}
\usepackage[mathscr]{euscript}
\usepackage{lipsum}
\usepackage{mathtools}
\usepackage{cuted}
\usepackage[export]{adjustbox}

\usepackage{subcaption}
\usepackage[colorlinks=true, urlcolor=blue, linkcolor=red]{hyperref}

\usepackage{mathtools}

\newcommand{\arXiv}[2]{\href{http://arxiv.org/pdf/hep-th/#1}{{\tt #2/#1}}}

\newcommand{\arXivold}[2]{\href{http://arxiv.org/pdf/#1}{{\tt #2/#1}}}

\usepackage{blindtext} 

\begin{document}

\preprint{APS/123-QED}

\title{The Generalized Scalar Weak Gravity Conjecture and its Implications}
\author{Fayez Abu-Ajamieh}
 \email{fayezajamieh@iisc.ac.in}
\affiliation{%
Centre for High Energy Physics, Indian Institute of Science, Bangalore 560012, India
}
\author{Nobuchika Okada}
\email{okadan@ua.edu}
\affiliation{
Department of Physics and Astronomy; 
University of Alabama; Tuscaloosa; Alabama 35487; USA}

\author{Sudhir K. Vempati}
\email{vempati@iisc.ac.in}
\affiliation{Centre for High Energy Physics, Indian Institute of Science, Bangalore 560012, India}


\begin{abstract}
We propose a generalized formulation of the Scalar Weak Gravity Conjecture (SWGC) based on  the analogy with the derivation of the Gauge Weak Gravity Conjecture (GWGC). We discuss some phenomenological implications of this Generalized SWGC, including the scale of New Physics (NP) when applied to the SM Higgs sector, and the bounds on the axion's couplings to fermions and the photon when applied to the axion. The Generalized SWGC constraints rule out most of the parameter space of axion-nucleon couplings, leaving only a tiny parameter space.
\end{abstract}
\pacs{Valid PACS appear here}
\maketitle


\section{Introduction}\label{sec1}
Although gravity is the most familiar interaction that we experience on a day-to-day basis, it remains the least understood out of all the fundamental interactions. Quantum theories for the electromagnetic, strong and weak interactions exist and have proven successful, as evidenced by the many successes of the Standard Model (SM). However, a quantum theory for gravity is still lacking in spite of the existence of several candidates, most notable of which is string theory.

The swampland program \cite{Vafa:2005ui,Ooguri:2006in} attempts at qualifying Quantum Field Theories (QFTs) that are consistent with quantum gravity represented by string theory. According to the Swampland Conjecture, QFTs that can be successfully UV-completed to include quantum gravity are said to belong to the landscape, whereas those that cannot are said to belong to the swampland. Only the former QFTs are considered candidates for quantum gravity.

One of the main principles of the swampland program is that gravity should be the weakest force, as embodied by the GWGC \cite{Arkani-Hamed:2006emk}. The conjecture states that for any $U(1)$ gauge group with charge $q$, there must exist a particle of mass $m$ such that 
\begin{equation}\label{eq:Gauge WGC}
m \leq q M_{\text{Pl}},
\end{equation}
where $M_{\text{Pl}} = 2.4 \times 10^{18}$ GeV is the reduced Planck scale. The main motivation for this conjecture arises from extremal black holes: If a particle satisfying eq. (\ref{eq:Gauge WGC}) does not exist, then charged black holes cannot evaporate, and we are faced with the problem of remnants \cite{Susskind:1995da}. 

Eq. (\ref{eq:Gauge WGC}) is usually called the electric WGC and should also apply to magnetic monopoles
\begin{equation}\label{eq:Mag_monopole}
m_{\text{mag}} \lesssim g_{\text{mag}} M_{\text{Pl}} \sim \frac{1}{q} M_{\text{Pl}}.
\end{equation}

As monopoles have mass that is at least of the order of the magnetic field it generates, which is linearly divergent, one has
\begin{equation}\label{eq:Monopole_mass}
m_{\text{mag}} \sim \frac{\Lambda}{q^{2}}.
\end{equation}
Plugging eq. (\ref{eq:Monopole_mass}) in eq. (\ref{eq:Mag_monopole}), we find
\begin{equation}\label{eq:Magnetic_WGC}
\Lambda \lesssim q M_{\text{Pl}}.
\end{equation}
Eq. (\ref{eq:Magnetic_WGC}) is called the magnetic WGC and it implies that there is a natural cutoff for any $U(1)$ gauge theory where the Effective Field Theory (EFT) breaks down.\footnote{The magnetic WGC was challenged in \cite{Saraswat:2016eaz}, although the electric form was argued to continue to hold in the same reference.}

The natural question that arises is: What about scalar interactions? In other words, is gravity a weaker force than that generated by scalar interactions? There have been several attempts at formulating a Scalar WGC \cite{Lust:2017wrl, Gonzalo:2019gjp, Shirai:2019tgr, Heidenreich:2019zkl, Freivogel:2019mtr, Benakli:2020pkm}. In this \textit{letter}, we attempt at formulating a generalized form of the SWGC and we investigate some of its phenomenological implications.

This paper is organized as follows: In Section \ref{sec2} we review the several attempts at formulating a SWGC. In Section \ref{sec3} we present our Generalized SWGC. In Section \ref{sec4} we discuss some of its phenomenological implications and then we conclude in Section \ref{sec5}.

\section{The Different Forms of the SWGC}\label{sec2}
The first attempt at answering this question came in \cite{Palti:2017elp, Lust:2017wrl}. There, it was conjectured that the particle with the largest charge to mass ratio should not form gravitationally bound states, which implies that
\begin{equation}\label{eq:SWGC_1}
q^{2} \geq \frac{m^{2}}{M^{2}_{\text{Pl}}} + \sum_{i,j} g^{ij} (\partial_{\varphi_{i}} m)( \partial_{\varphi_{j}} m),
\end{equation}
where $g^{ij}$ is the metric in the field space of $\varphi_{i}$, which are the scalar mediators of the force between WGC particles of mass $m$. The mediating scalars $\varphi_{i}$ should be massless, as massive scalars will develop a Yukawa potential exponentially suppressed by its mass ($\sim e^{-m r}/r$), which means that at sufficiently large distances, the corresponding force will eventually becomes weaker than the gravitational force.

If the WGC particle has no charge, then one can recast eq. (\ref{eq:SWGC_1}) into a statement about the \textit{magnitude} of the forces
\begin{equation}\label{eq:SWGC_2}
|\partial_{\varphi} m|^{2} \equiv \sum_{i,j} g^{ij} (\partial_{\varphi_{i}}m)( \partial_{\varphi_{j}} m) \geq \frac{m^{2}}{M^{2}_{\text{Pl}}}.
\end{equation}
The physical content of eq. (\ref{eq:SWGC_2}) is that gravity is a weaker force than that mediated by a (massless) scalar force
\begin{equation}\label{eq:SWGC_3}
F_{\text{garv}} = \frac{m^{2}}{8\pi M_{\text{Pl}}^{2}r^{2}} \leq F_{\text{scalar}} = \frac{|\partial_{\varphi} m|^{2}}{4\pi r^{2}}.
\end{equation}

The masslessness assumption of $\varphi$ was relaxed in \cite{Shirai:2019tgr} into $m_{\varphi} \lesssim 10^{-33}$ eV equivalent to the Hubble radius, which means that gravity should be weaker than a scalar force within the observable universe. Also, \cite{Shirai:2019tgr} showed that the above form of the SWGC was in tension with fifth force searches and with the de Sitter Swampland Conjecture \cite{Obied:2018sgi}.

In \cite{Gonzalo:2019gjp} a Strong SWGC was introduced. There, it was stated that the potential of any canonically normalized real scalar field must satisfy the constraint
\begin{equation}\label{eq:Strong_SWCG}
2(V''')^{2} - V''V'''' \geq \frac{(V'')^{2}}{M_{Pl}^{2}},
\end{equation} 
where the primes indicate derivatives with respect to the field. The above form was motivated by the desire to extend the original SWGC to all scalar fields (and not just scalar field whose masses are functions of $\varphi$), and to accommodate the periodic potential of axions. However, a (possible) counterexample was proposed in \cite{Freivogel:2019mtr}.

\cite{Heidenreich:2019zkl} introduced the Repulsive Force Conjecture (RFC), a refinement of the the original SWGC in eq. (\ref{eq:SWGC_1}) which states that the force between two copies of a charged particle must be repulsive. In other words, since gravitational and scalar interactions are both attractive, the repulsive $U(1)$ force should be larger than both forces combined, essentially leading to eq. (\ref{eq:SWGC_1}).

Another refinement of the SWGC was introduced in \cite{Benakli:2020pkm}, where it was postulated that in the low energy (non-relativistic) limit, the gravitational contribution to the leading interaction must be subleading. The conjecture was applied to a \textit{massive} $\phi^{4}$ theory, where it was found that the SWGC can be violated in a small region of size $\frac{\Delta \phi^{2}}{m^{2}} \sim \frac{m^{2}}{M_{\text{Pl}}^{2}}$.

We should point out that all attempts to formulate a SWGC stand on a weaker footing compared to the GWGC due to the lack of a black hole evaporation argument in the scalar case. Therefore, all forms of the SWGC should be considered with this caveat in mind.

\section{The Generalized SWGC}\label{sec3}
\begin{figure}[!t] 
\centering
\includegraphics[width=0.4\textwidth]{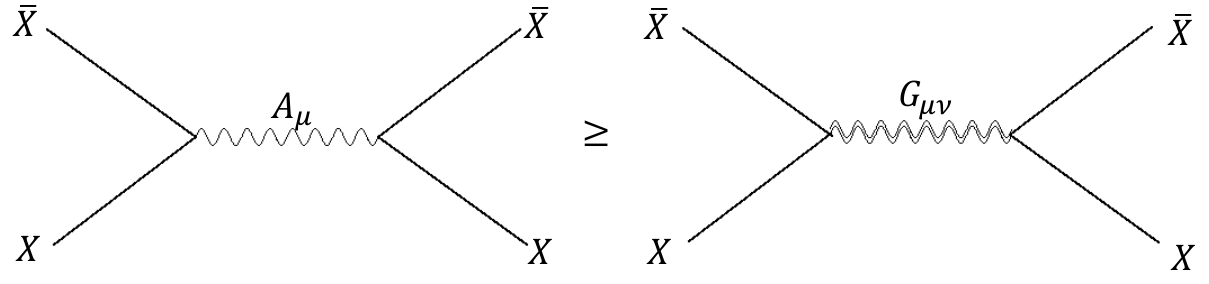}
\caption{Extracting the GWGC.}
\label{fig1}
\end{figure}

Let's first consider the magnetic GWGC in eq. (\ref{eq:Magnetic_WGC}). Notice that this bound can be obtained by requiring that gravity be a weaker \textit{interaction} than that of a gauged $U(1)$. More specifically, if we require gravitational interactions to be always weaker than $U(1)$ interactions at \textit{any energy} and not just in the non-relativistic limit (see Fig. \ref{fig1}) and impose
\begin{equation}\label{eq:general_GWGC}
|\mathcal{M}_{\text{grav}}(s)| \lesssim |\mathcal{M}_{\text{U(1)}}(s)|,
\end{equation}
then we can see that in the limit $\sqrt{s} \gg m$, where $m$ is the mass of the WGC particle,  eq. (\ref{eq:general_GWGC}) implies that $q^{2} \lesssim \frac{s}{M_{Pl}^{2}}$. Setting $\sqrt{s} = \Lambda$ as the scale of NP, we retrieve eq. (\ref{eq:Magnetic_WGC}). 

One can also argue that even if the charge carriers inside black holes have high energy, black holes should still be able to evaporate, and thus the electric WGC should also extend beyond the non-relativistic regime. This motivates us to generalize the SWGC beyond the non-relativistic regime by requiring that 
\begin{equation}\label{eq:general_SWGC}
|\mathcal{M}_{\text{grav}}(s)| \lesssim |\mathcal{M}_{\text{scalar}}(s)|.
\end{equation}

In applying eq. (\ref{eq:general_SWGC}), we should keep in mind that massless gravity suffers from a divergence in the $t$- and $u$- channels. Thus, to be conservative, we only consider the (finite) $s$- channel when evaluating the gravitational amplitude. Notice that unlike the original formulation of the SWGC in eqs. (\ref{eq:SWGC_1}) and (\ref{eq:SWGC_2}), the Generalized SWGC applies to both massless and massive scalar mediator and not just massless ones.

\section{Some Phenomenological Implications of the WGC}\label{sec4}
Beyond qualifying candidate theories for quantum gravity, both the GWGC and the SWGC have some interesting phenomenological implications. The implications of the GWGC were discussed in \cite{Abu-Ajamieh:2024gaw}. As shown there in detail, one can use the magnetic form of the GWGC to set a limit on the scale of NP that corresponds to a certain $U(1)$ gauge theory if the charge is known. For example, the cutoff scale that corresponds to $U(1)_{\text{EM}}$ is $\Lambda \sim 10^{17}$ GeV. Conversely, if a lower limit on the scale of NP is known (for example from null collider searchers), then the magnetic GWGC can be used to set a lower limit on the charge of that $U(1)$. 

A similar argument holds for the SWGC. Here, we discuss some pehomenological implications of the SWGC in two case: When the mediator is the SM Higgs boson and when the mediator is the axion. In our calculation, we limit ourselves to amplitudes at tree-level.
\subsection{The SM Higgs}
To the best of our knowledge, the SM Higgs is the only fundamental scalar that exists. If we apply the SWGC where the Higgs is the scalar mediator and the other SM particles are the WGC states, we can set limits on the scale of NP in each state.
\subsubsection{Fermion WGC States}
The Higgs-mediated $\overline{f}f \rightarrow \overline{f}f$ proceeds via the $s$- and $t$- channels. The amplitude reads
\begin{equation}\label{eq:ffHff_amp}
|\mathcal{M}_{h}|^{2} = y_{f}^{4} \frac{\Big(3(1+\tau +\tau^{2}) + (3 - 3 \tau +\tau^{2})\cos^{2}{\theta} -6 \cos{\theta}\Big)}{(1-\tau)^{2}(1 - \cos{\theta} + 2\tau)^{2}},
\end{equation}
where $\tau  = \frac{m_{h}^{2}}{s}$ and we have dropped the masses of the initial and final state fermions. On the other hand, the gravitational amplitude reads
\begin{equation}\label{eq:ffGff_amp}
|\mathcal{M}_{\text{grav}}|^{2} = \frac{s^{2}}{32M_{\text{Pl}}^{4}}(2 + \cos{(2\theta)}+\cos(4\theta)),
\end{equation}
Applying the Generalized SWGC in eq. (\ref{eq:general_SWGC}), it is easy to see that the strongest bound arises in the forward region $\theta \rightarrow 0$. In the high energy limit $\tau \rightarrow 0$, we have
\begin{equation}\label{eq:H_fermion_SWGC}
\frac{s}{2\sqrt{2}M^{2}_{\text{Pl}}} \lesssim y_{f}^{2}.
\end{equation}

\begin{figure}[t!]
\centering
\begin{subfigure}[b]{0.38\textwidth}
   \includegraphics[width=1\linewidth]{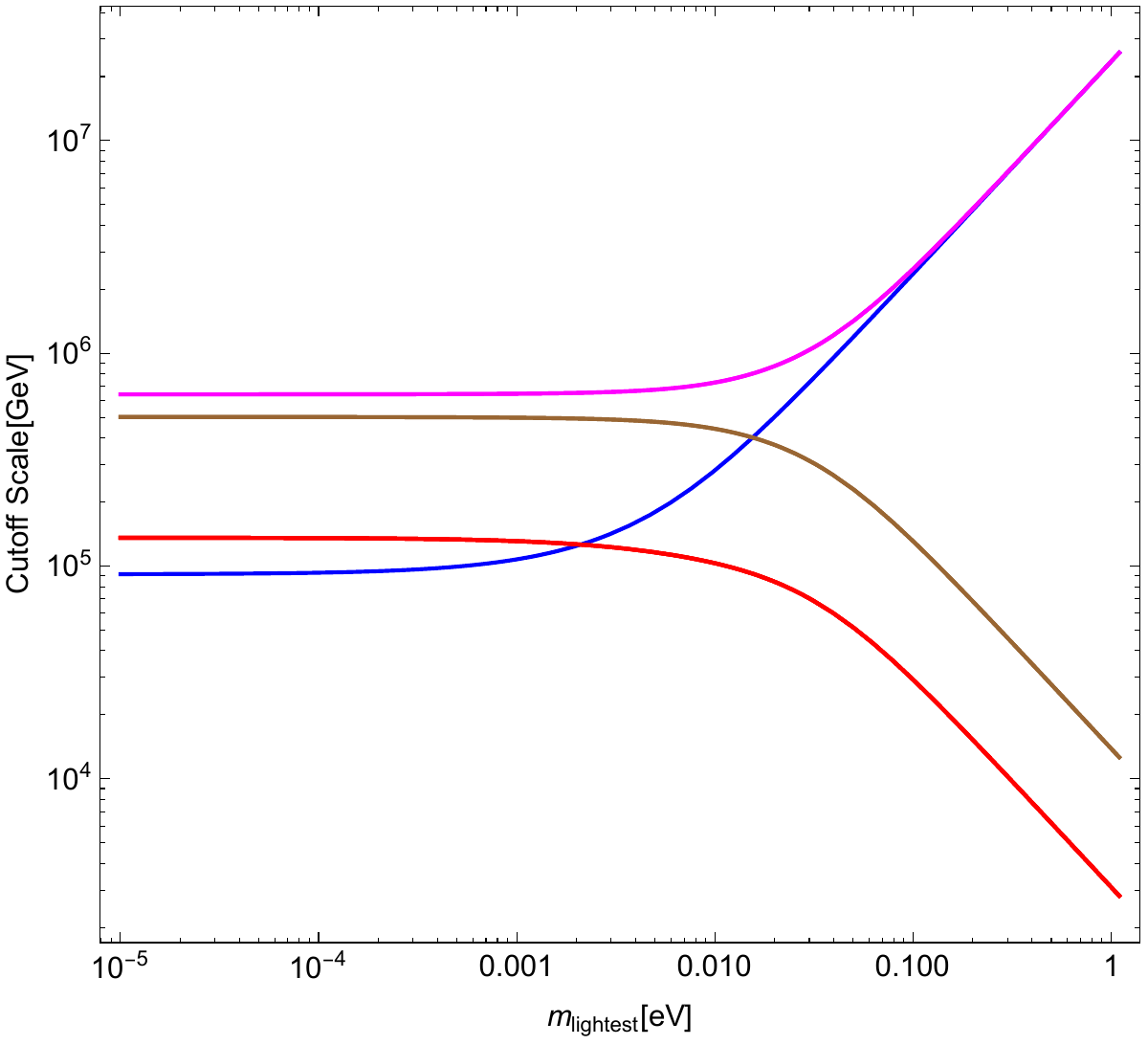} 
\end{subfigure}

\begin{subfigure}[b]{0.38\textwidth}
   \includegraphics[width=1\linewidth]{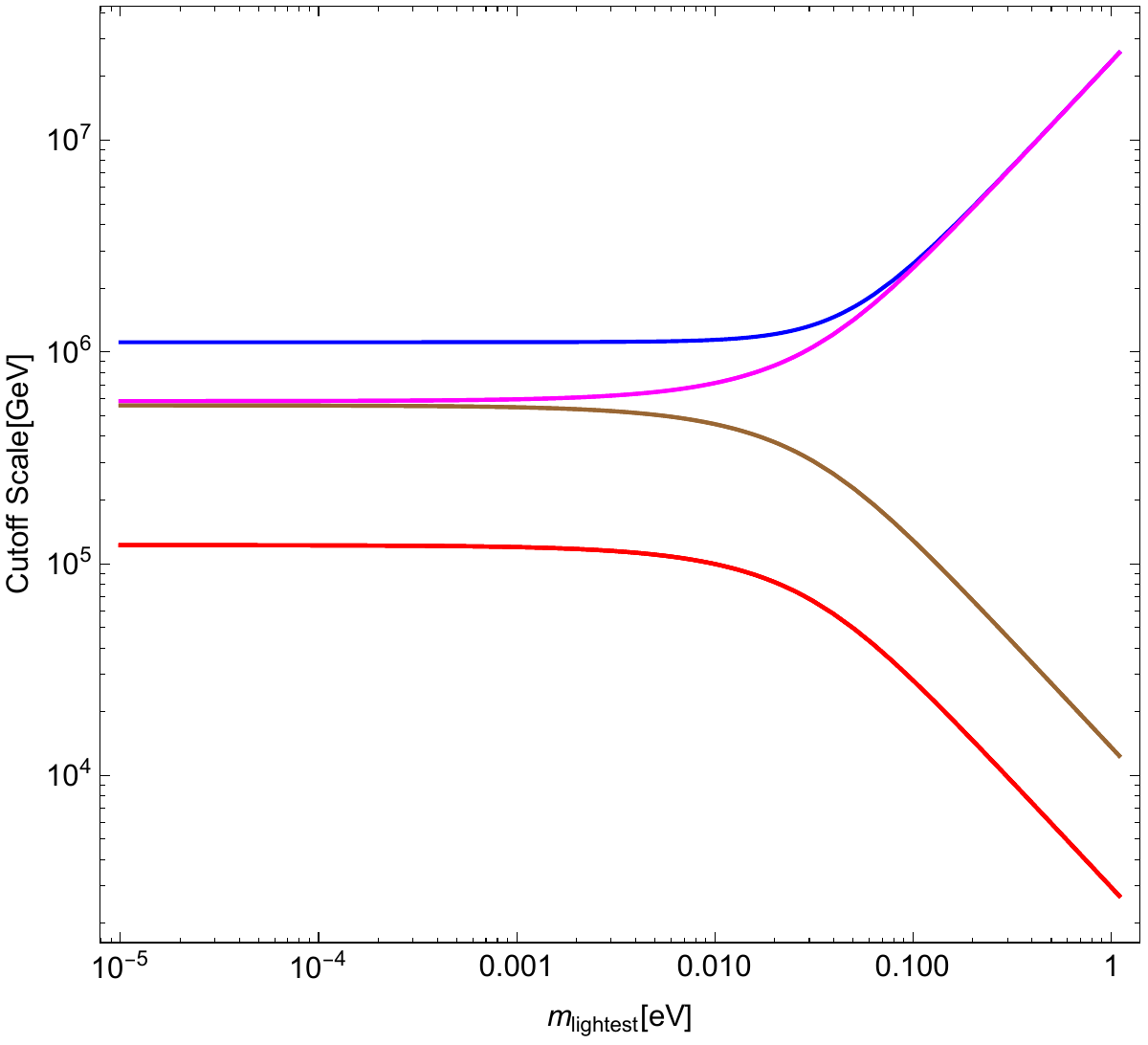}
\end{subfigure}
\caption{The scale of NP as a function of the lightest neutrino mass from neutrino oscillation data for the normal (top) and the inverted (bottom) hierarchies. The different lines correspond to the eigenstates of the neutrino mass matrix: The blue line represents $m_{11}$, the magenta is $m_{22} = m_{33}$, the brown is $m_{23} = m_{32}$, and the red line is $m_{12} = m_{21} = m_{13} = m_{31}$.}
\label{fig2}
\end{figure}

Setting $\sqrt{s} = \Lambda$ as the scale of NP, we find that $\Lambda \lesssim \sqrt[4]{8}y_{f}M_{\text{Pl}}$. Obviously the scale of NP is set by the lightest fermion. If neutrinos are Majorana fermions, then the electron will be the lightest SM fermion and $\Lambda \lesssim 1.2 \times 10^{13}$ GeV. On the other hand, if neutrinos are Dirac fermions, then the scale of NP is set by the mass of the lightest neutrino. If we naively use the upper limit on $m_{\nu_{e}} = 1.1$ eV as indicated by the PDG \cite{ParticleDataGroup:2020ssz}, then the scale on NP becomes $\Lambda \lesssim 2.6 \times 10^{7}$ GeV. However, much stronger bounds can be obtained using the neutrino oscillation data. Fig. \ref{fig2} shows the scale of NP as a function of the mass of the lightest neutrino for both the normal and the inverted hierarchies. Fig. \ref{fig2} was created using neutrino oscillation data as input to the neutrino mixing (PMNS) matrix, where the input values are $\sin^{2}{(2\theta_{12})} = 0.87$, $\sin^{2}{(2\theta_{23})} = 1.0$, $\sin^{2}{(2\theta_{13}) = 0.092}$, $\delta_{CP} = \frac{3}{2}\pi$, $\Delta m_{21}^{2} = 7.6 \times 10^{-5}$ $\text{eV}^{2}$ for both the normal and the inverted hierarchies, and  $|\Delta m_{31}^{2}(\Delta m_{32}^{2})| = 2.4 \times 10^{-3}$ $ \text{eV}^{2}$ for the normal (inverted hierarchy).

We can see from the plots that for the normal hierarchy, $\Lambda$ ranges between $\sim 2.8$ TeV and $\sim 92$ TeV corresponding to a mass of $1.1$ eV to $\lesssim 10^{-3}$ eV for the lightest neutrino respectively, whereas for the inverted heirarchy, it ranges between $\sim 2.7$ TeV and $\sim 122$ TeV for the same neutrino mass range. The lower range of this scale can be potentially probed in future colliders, such as the 100 TeV FCC or the muon collider, and can even be within the reach of the LHC.

\subsubsection{Massive Gauge Boson WGC States}
If we compare the amplitude of $WW (ZZ) \rightarrow WW (ZZ)$ mediated through the Higgs, with that mediated through a graviton in the high energy limit $\sqrt{s} \gg m_{W}, m_{Z}, m_{h}, v$, then in the forward region, the Generalized SWGC leads to the bound
\begin{equation}\label{eq:H_V_SWGC}
\Lambda \lesssim \frac{g_{V}^{2}vM_{\text{Pl}}}{\sqrt[4]{2}m_{h}},
\end{equation}
where $g_{W}^{2} = g^{2}$ and  $g_{Z}^{2} = g^{2}+g'^{2}$. This corresponds to $\Lambda \sim 5 \times 10^{17} (2.2 \times 10^{18})$ GeV for the $W(Z)$.

\subsubsection{Massless Gauge Boson WGC States}
Although the Higgs boson does not couple to photons and gluons at tree-level, we can nonetheless integrate out the triangle loops and write the effective coupling as
\begin{equation}\label{eq:Eff_lag}
\mathcal{L} = c_{\gamma} \frac{\alpha}{\pi v} h A_{\mu\nu}A^{\mu\nu} + c_{g} \frac{\alpha_{s}}{12\pi v} h G^{a}_{\mu\nu}G^{a\mu\nu},
\end{equation}
where for the SM, $c_{\gamma} \simeq 0.81$, $c_{g} \simeq 1.03$. Applying eq. (\ref{eq:general_SWGC}) to the photon sector in the forward region, we find
\begin{equation}\label{eq:H_A_SWGC}
\Lambda \lesssim \frac{2\sqrt[4]{8}\alpha c_{\gamma}}{v \pi} m_{h} M_{\text{Pl}} \simeq 7.8 \times 10^{15} \hspace{1mm} \text{GeV}.
\end{equation}

Things are a bit more subtle in the gluon sector, as we need to take into consideration the running of $\alpha_{s}$. Applying the Generalized SWGC, we find
\begin{equation}\label{eq:eq:H_G_SWGC}
\Lambda \lesssim \frac{c_{g} m_{h} M_{\text{Pl}}}{3\sqrt[4]{2}\pi v} \frac{\alpha_{s}(M_{Z})}{1-\frac{b_{0} \alpha_{s}(M_{Z})}{2\pi}\log(\frac{\Lambda}{M_{Z}})} \simeq 2.6 \times 10^{15} \hspace{1mm} \text{GeV},
\end{equation}
where $b_{0} = -11 +\frac{2}{3}n_{f} = -7$ and $\alpha_{s}(M_{Z}) = 0.1179$.

\subsubsection{Scalar WGC States}
Finally, let's consider the case where the Higgs is both the mediator and the WGC particle. In this case, $hh \rightarrow hh$ proceeds via the $s$-, $t$- and $u$- channels in addition to a contact term via the quartic coupling. In the high energy limit, the quartic term dominates and Generalized SWGC implies 
\begin{equation}\label{eq:eq:H_H_SWGC}
\frac{s \sin^{2}{\theta}}{4 M_{\text{P}}^{2}} \lesssim 6\lambda \hspace{2mm} \implies   \hspace{2mm} \lambda \gtrsim \frac{\Lambda^{2}}{24M_{\text{P}}^{2}}.
\end{equation}
where we took $\theta \rightarrow \frac{\pi}{2}$ as it yields the strongest bound. As it is well-known, in the SM, the Higgs sector suffers from an instability when the quartic coupling runs into negative values at a scale $\sim 10^{10} - 10^{11}$ GeV, however, the Generalized SWGC suggests that the quartic coupling must remain positive, thereby avoiding any instability. What this means is that the Generalized SWGC (if indeed correct), informs us that there must be NP at an energy scale before the supposed scale of instability, that ensures the stability of the Higgs potential.

\subsection{Axions}
We now consider the axion as the mediating particle, and cover the cases where the WGC states are fermions and photons. The axion interaction Lagrangian is given by
\begin{equation}\label{eq:Axion_lag}
\mathcal{L}_{\text{int}} = -\frac{1}{f} g_{a\gamma} a F_{\mu\nu}\widetilde{F}^{\mu\nu} + \frac{g_{af}}{2m_{f}}(\partial_{\mu}a)(\overline{f}\gamma^{\mu}\gamma^{5}f),
\end{equation}
where $\widetilde{F}^{\mu\nu} = \frac{1}{2}\epsilon^{\mu\nu\rho\sigma}F_{\rho\sigma}$. We should point out that since the axion corresponds to a $U(1)$ gauge group, the bounds in this section can also be obtained through the GWGC (see \cite{Abu-Ajamieh:2024gaw}), however, here we are more rigorous.

\subsubsection{Fermion WGC States}
First, let's consider the case $\overline{f}f \rightarrow \overline{\psi}\psi$ where $f \neq \psi$. In this case, the axion-mediated process proceeds through the $s$-channel only. A simple calculation shows that in the forward region, the Generalized SWGC (or Guage) implies
\begin{equation}\label{eq:axion_f_SWGC_1}
g_{af}g_{a\psi} \gtrsim \frac{|s-m_{a}^{2}|}{2\sqrt{2}M_{\text{Pl}}^{2}}.
\end{equation} 

\begin{figure}[t!] 
\centering
\includegraphics[width=0.45\textwidth]{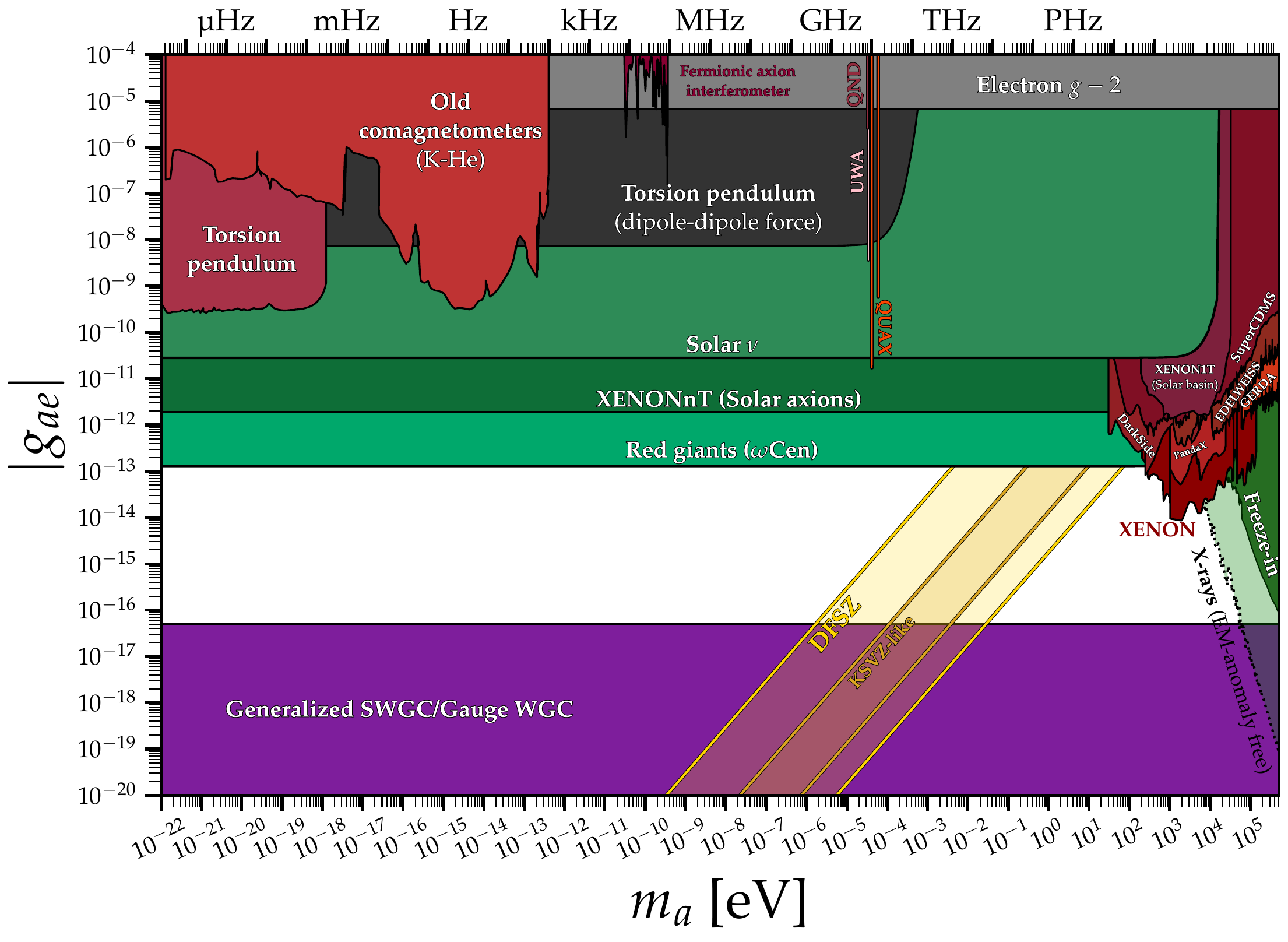}
\caption{The lower bound on the axion's coupling to electrons as suggested by the Gauge/Generalized Scalar WGC. The plot was modified from \cite{AxionLimits}.}
\label{fig3}
\end{figure}

On the other hand, if $f = \psi$, then there is a $t$-channel in addition to the $s$-channel, and $g_{af} = g_{a\psi}$. In this case, the conjecture becomes
\begin{equation}\label{eq:axion_f_SWGC_2}
g_{af} \gtrsim \frac{1}{M_{\text{Pl}}} \sqrt{\frac{|s-m_{a}^{2}|}{2\sqrt{2}}}.
\end{equation}

Eqs. (\ref{eq:axion_f_SWGC_1}) and (\ref{eq:axion_f_SWGC_2}) can be used to set bounds on the axion's coupling to fermions, however, to do that, we need a lower limit on the scale of NP. As explained in \cite{Abu-Ajamieh:2024gaw}, LEP searches $e^{+}e^{-} \rightarrow a \rightarrow \bar{f}f$ can be used, and the lower limit on the NP can be set as $\Lambda = \sqrt{s_{\text{LEP}}} = 209$ GeV. In this case, eq.~(\ref{eq:axion_f_SWGC_2}) can be used to set a lower bound on $g_{ae}$. Using the resources from \cite{AxionLimits}, we show in Fig. \ref{fig3} the lower bound on $g_{ae}$ as suggested by the WGC (scalar or gauge), superimposed on the other experimental bounds. As can seen from the plot, while experimental bounds place limits on the coupling from above, the Scalar/Gauge WGC places limits from below, and thus can help probe the entire parameter space.

\begin{figure}[!t]
\centering
\begin{subfigure}[b]{0.45\textwidth}
   \includegraphics[width=1\linewidth]{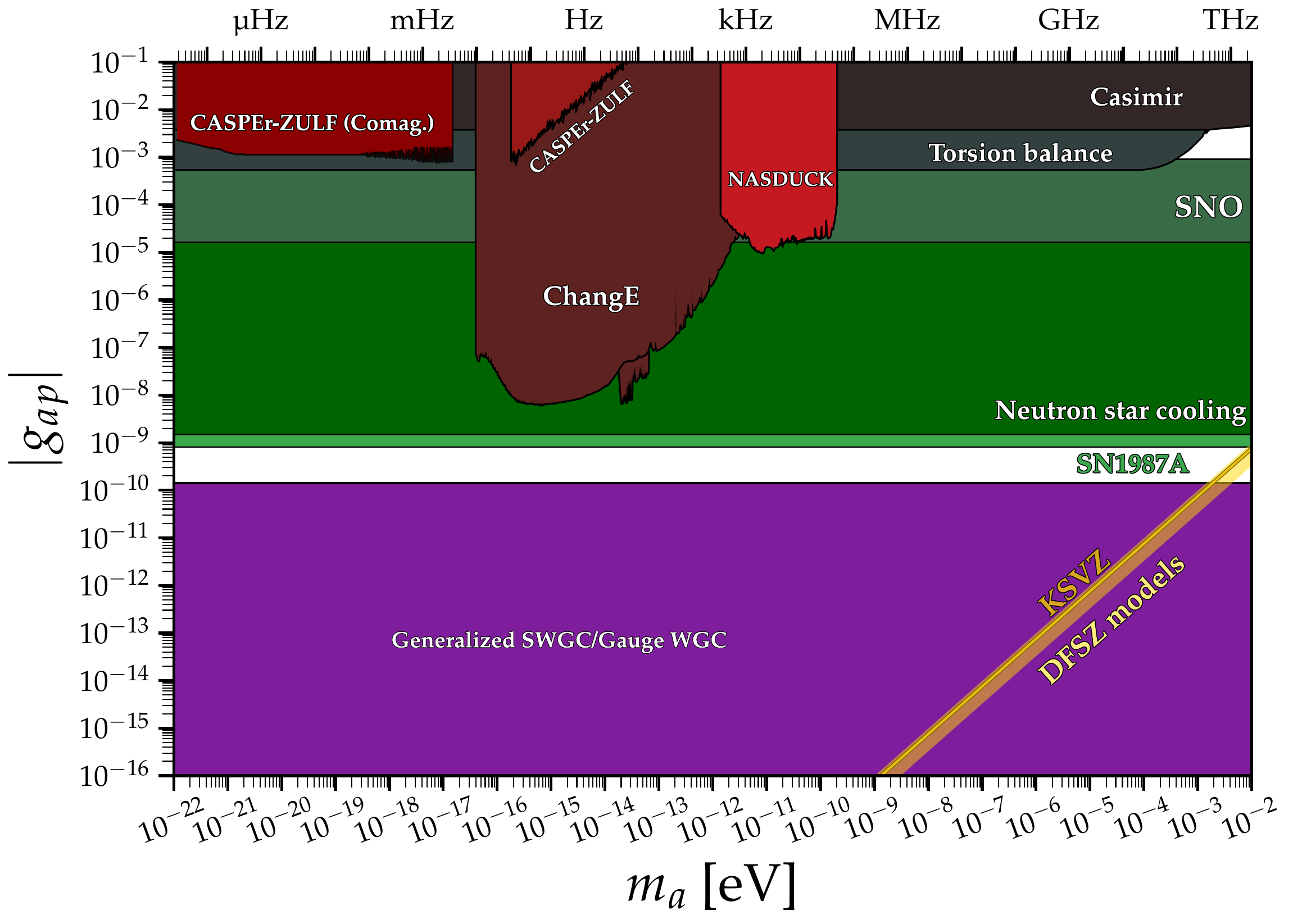} 
\end{subfigure}

\begin{subfigure}[b]{0.45\textwidth}
   \includegraphics[width=1\linewidth]{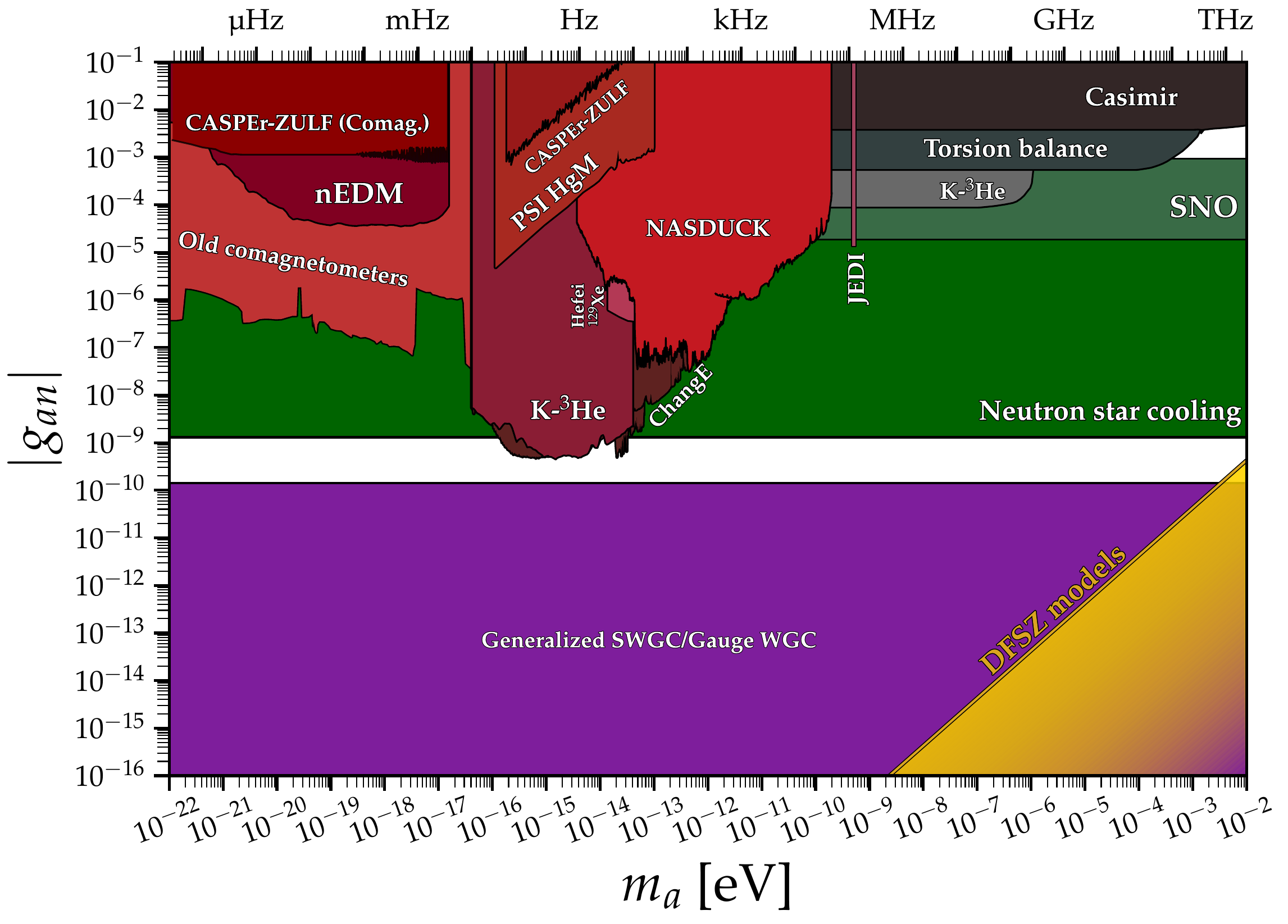}
\end{subfigure}
\caption{The lower bound on the axion's coupling to protons (up) and neutrons (down) as suggested by the Gauge/Generalized Scalar WGC. The plots were modified from \cite{AxionLimits}.}
\label{fig4}
\end{figure}

On the other hand, to set bounds on the axion couplings to other fermions, we need to use eq. (\ref{eq:axion_f_SWGC_1}). To do so, we need to specify $g_{ae}$. To set conservative bounds, we set $g_{ae}$ to be equal to the lower experimental bounds extracted from Fig. \ref{fig3}. Fig. \ref{fig4} shows the bounds on the axion coupling to protons and neutrons. As the plots show, the Scalar/Gauge WGC can place very stringent constraints on the parameter space, with only a small window remaining open in both cases. The bounds on the axion couplings to other fermions are identical to those to protons and neutrons as can be seen from eq. (\ref{eq:axion_f_SWGC_2}).

\subsubsection{Photon WGC}
Next we consider $\gamma\gamma \rightarrow a \rightarrow  \gamma\gamma$. This process proceeds via the $s$-, $t$-, and $u$-channels. Applying the Generalized SWGC leads to the bound
\begin{equation}\label{eq:axion_A_SWGC}
g_{\gamma a} \gtrsim \frac{1}{M_{\text{Pl}}} \Bigg( \frac{8(1-m_{a}^{4}/s^{2})^{2}}{1+3m_{a}^{4}/s^{2}}\Bigg)^{\frac{1}{4}}.
\end{equation}

\begin{figure}[!h] 
\centering
\includegraphics[width=0.45\textwidth]{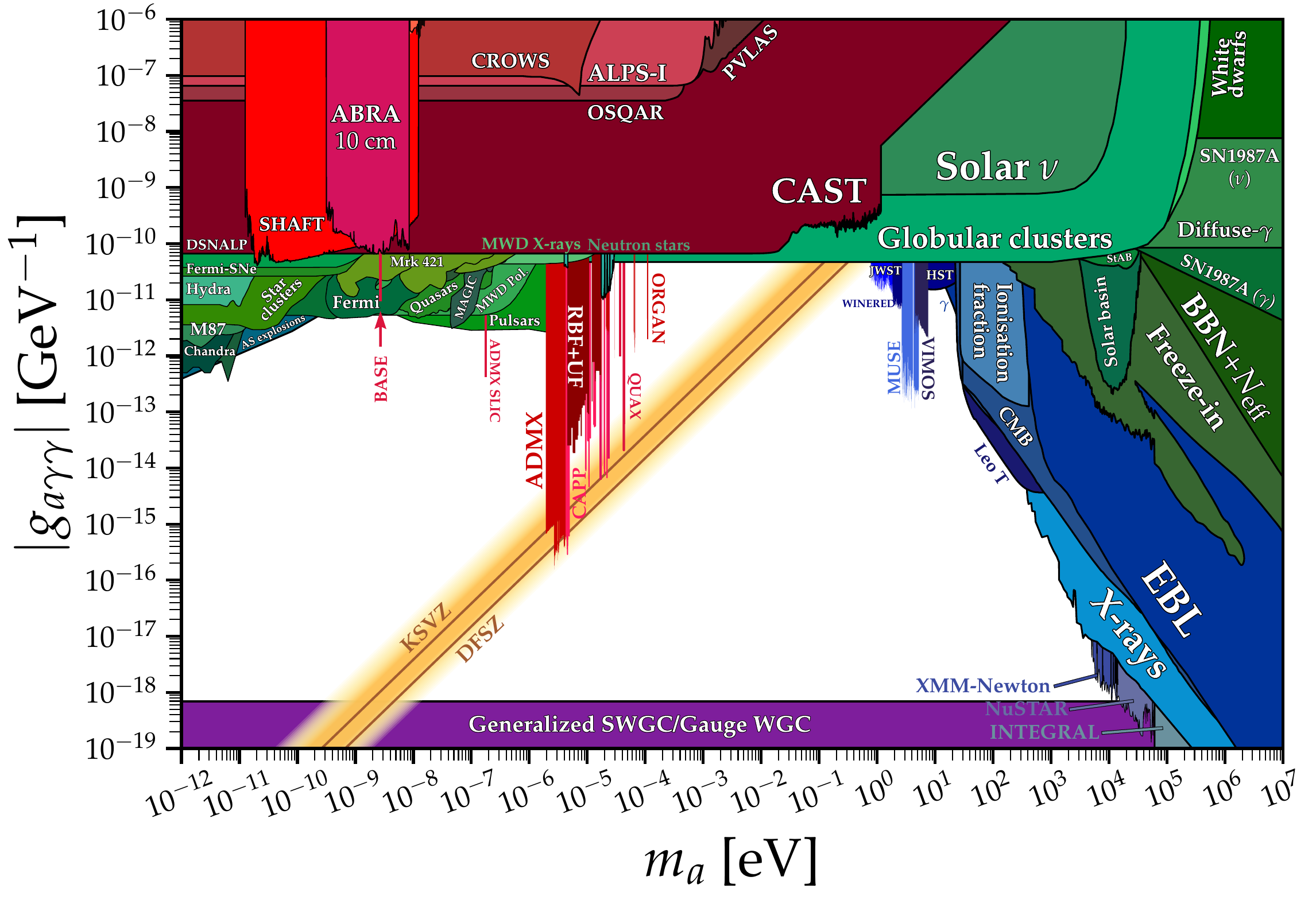}
\caption{The lower bound on the axion's coupling to photons as suggested by the Gauge/Generalized Scalar WGC. The plot was modified from \cite{AxionLimits}.}
\label{fig5}
\end{figure}

Here too, we need to a lower limit on the scale of NP to be able to set bounds on $g_{a\gamma}$. As explained in detail in \cite{Abu-Ajamieh:2024gaw}, the latest Atlas results on the Light-by-Light (LBL) scattering \cite{ATLAS:2017fur} can be used to infer a lower bound on the scale of NP. Following \cite{Ellis:2017edi, Ellis:2022uxv}, it was found in \cite{Abu-Ajamieh:2024gaw} that $\sqrt{s}\equiv \Lambda_{\text{LBL}} \simeq 7.8$ TeV can be obtained. Using this in eq. (\ref{eq:axion_A_SWGC}), we plot the bound on the axion's coupling to photons in Fig. \ref{fig5}, where we can see from the plot that compared to the coupling to fermions, the bound is weaker. This is due to the energy behavior of the amplitudes. While at high energy, the amplitude of the fermionic WGC states becomes essentially constant, the amplitude of the photon WGC states grows quadratically with energy, which weakens the bound.

\section{Conclusions}\label{sec5}
In this paper, we proposed a generalized formulation of the SWGC, where we suggested that gravity is always a weaker interaction than scalar interactions at any energy scale, whether the scalar is massless or massive. We proceeded to investigate some of the phenomenological implications of the Generalized SWGC when the scalar is the SM Higgs and the axion. In the former case, we found that the Generalized SWGC can set an upper bound on the scale of NP $\sim 10^{13}$ GeV if neutrinos are Majorana fermions, and as low as $\sim 3 - 122$ TeV if neutrinos are Dirac fermions. We also showed that the Generalized SWGC suggests that the Higgs quartic coupling should always remain positive, indicating the absence of any instability in the Higgs potential.

We showed that both the Generalized SWGC and GWGC both set stringent lower bounds on the axion couplings to fermions and to the photon, excluding much of the parameter space. We should emphasize that due to the lack of a black hole evaporation argument, the SWGC with its various formations, stands on weaker grounds compared to the GWGC.

\section*{Acknowledgments}
The work of NO is supported in part by the United States Department of Energy (DC-SC 0012447 and DC-SC 0023713). 
SKV is supported by SERB, DST, Govt. of India  Grants MTR/2022/000255 , “Theoretical aspects of some physics beyond standard models”,  
CRG/2021/007170 “Tiny Effects from Heavy New Physics “and IoE funds from IISC.

\appendix


\begin{thebibliography}{10}
 
\bibitem{Vafa:2005ui}
C.~Vafa,
``The String landscape and the swampland,''
arXivold{hep-th/0509212}{hep-th}.


\bibitem{Ooguri:2006in}
H.~Ooguri and C.~Vafa,
``On the Geometry of the String Landscape and the Swampland,''
Nucl. Phys. B \textbf{766}, 21-33 (2007)
\arXivold{hep-th/0605264}{hep-th}.

\bibitem{Arkani-Hamed:2006emk}
N.~Arkani-Hamed, L.~Motl, A.~Nicolis and C.~Vafa,
``The String landscape, black holes and gravity as the weakest force,''
JHEP \textbf{06}, 060 (2007)
\arXiv{0601001}{hep-th}.

\bibitem{Susskind:1995da}
L.~Susskind,
``Trouble for remnants,''
\arXiv{9501106}{hep-th}.

\bibitem{Saraswat:2016eaz}
P.~Saraswat,
``Weak gravity conjecture and effective field theory,''
Phys. Rev. D \textbf{95}, no.2, 025013 (2017)
\arXivold{1608.06951}{hep-th}.

\bibitem{Palti:2017elp}
E.~Palti,
``The Weak Gravity Conjecture and Scalar Fields,''
JHEP \textbf{08}, 034 (2017)
\arXivold{1705.04328}{hep-th}.

\bibitem{Lust:2017wrl}
D.~Lust and E.~Palti,
``Scalar Fields, Hierarchical UV/IR Mixing and The Weak Gravity Conjecture,''
JHEP \textbf{02}, 040 (2018)
\arXivold{1709.01790}{hep-th}.

\bibitem{Shirai:2019tgr}
S.~Shirai and M.~Yamazaki,
``Is Gravity the Weakest Force?,''
Class. Quant. Grav. \textbf{38}, no.3, 035006 (2021)
\arXivold{1904.10577}{hep-th}.


\bibitem{Gonzalo:2019gjp}
E.~Gonzalo and L.~E.~Ib\'a\~nez,
``A Strong Scalar Weak Gravity Conjecture and Some Implications,''
JHEP \textbf{08}, 118 (2019)
\arXivold{1903.08878}{hep-th}.

\bibitem{Freivogel:2019mtr}
B.~Freivogel, T.~Gasenzer, A.~Hebecker and S.~Leonhardt,
``A Conjecture on the Minimal Size of Bound States,''
SciPost Phys. \textbf{8}, no.4, 058 (2020)
\arXivold{1912.09485}{hep-th}.

\bibitem{Heidenreich:2019zkl}
B.~Heidenreich, M.~Reece and T.~Rudelius,
``Repulsive Forces and the Weak Gravity Conjecture,''
JHEP \textbf{10}, 055 (2019)
\arXivold{1906.02206}{hep-th}.

\bibitem{Benakli:2020pkm}
K.~Benakli, C.~Branchina and G.~Lafforgue-Marmet,
``Revisiting the scalar weak gravity conjecture,''
Eur. Phys. J. C \textbf{80}, no.8, 742 (2020)
\arXivold{2004.12476}{hep-th}.

\bibitem{Obied:2018sgi}
G.~Obied, H.~Ooguri, L.~Spodyneiko and C.~Vafa,
``De Sitter Space and the Swampland,''
\arXivold{1806.08362}{hep-th}.


\bibitem{Abu-Ajamieh:2024gaw}
F.~Abu-Ajamieh, N.~Okada and S.~K.~Vempati,
``Implications of the Weak Gravity Conjecture on Charge, Kinetic Mixing, the Photon Mass, and More,''
\arXivold{2401.10792}{hep-ph}.


\bibitem{ParticleDataGroup:2020ssz}
P.~A.~Zyla \textit{et al.} [Particle Data Group],
``Review of Particle Physics,''
PTEP \textbf{2020}, no.8, 083C01 (2020)

\bibitem{AxionLimits}
Ciaran O'Hare,
https://cajohare.github.io/AxionLimits/
\href{https://cajohare.github.io/AxionLimits/}{Axion Limits}

\bibitem{ATLAS:2017fur}
M.~Aaboud \textit{et al.} [ATLAS],
``Evidence for light-by-light scattering in heavy-ion collisions with the ATLAS detector at the LHC,''
Nature Phys. \textbf{13}, no.9, 852-858 (2017)
\arXivold{1702.01625}{hep-ex}.


\bibitem{Ellis:2017edi}
J.~Ellis, N.~E.~Mavromatos and T.~You,
``Light-by-Light Scattering Constraint on Born-Infeld Theory,''
Phys. Rev. Lett. \textbf{118}, no.26, 261802 (2017)
\arXivold{1703.08450}{hep-ph}.

\bibitem{Ellis:2022uxv}
J.~Ellis, N.~E.~Mavromatos, P.~Roloff and T.~You,
``Light-by-light scattering at future $e^+e^-$ colliders,''
Eur. Phys. J. C \textbf{82}, no.7, 634 (2022)
\arXivold{2203.17111}{hep-ph}.



\end{thebibliography}
\end{document}